\documentclass[aps,pra,twocolumn]{revtex4-1}

\usepackage{graphicx}

\def\qr{{\bf r}}                                   
\def\qk{{\bf k}}                                   
\def\qq{{\bf q}}                                   

\begin{document}

\title{Stability and excitations of a bilayer of strongly correlated dipolar Bosons}

\author{D. Hufnagl$^{1,2}$ and R. E. Zillich$^{1,3,4}$}

\affiliation{$^1$Institute for Theoretical Physics, Johannes Kepler
  University, Altenbergerstrasse 69, 4040 Linz, Austria\\
  $^2$Johann Radon Institute for Computational and Applied Mathematics (RICAM), Austrian Academy of Sciences,
Altenbergerstr. 69, 4040 Linz, Austria\\
  $^3$Institute for Quantum Optics and Quantum Information (IQOQI), Austrian Academy of Sciences, Otto Hittmair-Platz 1, 6020 Innsbruck, Austria\\
  $^4$Kavli Institute for Theoretical Physics (KITP), University of California, Santa Barbara CA 93106, USA}

\begin{abstract}

We study correlation effects and excitations in a dipolar Bose gas bilayer which is modeled by a
one-dimensional double well trap that determines the width of an individual layer,
the distance between the two layers, and the height of the barrier between them.
For the ground state calculations we use the hypernetted--chain Euler Lagrange
method and for the calculation 
of the excitations we use the correlated basis function method.  We observe
instabilities both for wide, well-separated layers dominated by \emph{intra-layer}
attraction of the dipoles, and for narrow layers that are close to each other dominated
by \emph{inter-layer} attraction.  The behavior of the pair distribution function
leads to the interpretation that the monomer phase becomes unstable when pairing
of two dipoles becomes energetically favorable between or within layers, respectively.
In both cases we observe a tendency towards ``rotonization'', i.e. the appearance of a soft mode
with finite momentum in the excitation spectrum.
The dynamic structure function is not simply characterized by a single excitation mode, but
has a non-trivial multi-peak structure that is not captured by the Bijl-Feynman
approximation.  The dipole-dipole interaction between different layers
leads to additional damping compared to the damping obtained for uncoupled layers.

\end{abstract}



\maketitle

\section{Introduction}

Experimental advances in achieving Bose-Einstein condensation (BEC) of atoms with large
magnetic moment ($^{52}$Cr~\cite{lahayeNature07,lahayePRL08},
$^{164}$Dy~\cite{luPRL11dysprosium}, $^{168}$Er,~\cite{aikawaPRL12})
and in generating quantum gases of heteronuclear molecules
(KRb~\cite{niScience08,ospelkausNatPhys08,niPCCP09,niNature10},
LiCs~\cite{deiglmayrPRL08},
LiK~\cite{voigtPRL09},
RbCs~\cite{sagePRL05,takekoshiPRA12})
by Feshbach association 
have lead to a growing interest in effects caused by the dipole-dipole interaction (DDI).
The shape of a trapped dipolar condensate~\cite{goralPRA00,ronenPRL07} and its stability against
collapse~\cite{lahayePRL08,kochNaturePhys08} have been investigated.
The dynamics~\cite{odellPRL03,santosPRL03,dipolePRL09,hufnaglPRL11,maciaPRL12} has been
studied theoretically, but recently also experimentally~\cite{bismutPRL12}.
The generation of novel phases with topological order using polar molecules has been
proposed~\cite{micheliNaturePhys06}.
A recent review of the field can be found in Ref.~\onlinecite{baranovChemPhys12}.
The strength of the DDI can be characterized by
the dipolar length $r_0=mC_{dd}/(4\pi\hbar^2)$, where $m$ is the mass of the dipolar
atom or molecule and $C_{dd}$ is proportional to the square of the dipole moment.
For magnetic moments, $r_0$ is usually much smaller than for electric dipole moments
of molecules, which can range to thousands of \AA .
Since achieving BEC with heteronuclear molecules is much harder than with
homonuclear molecules, Er$_2$ is a promising candiate to reach a much stronger DDI
regime with purely magnetic dipole moments~\cite{ferlainoprivatecomm}.

The two-dimensional limit of a dipolar Bose gas (DBG) polarized perpendicularly
to the plane has been studied extensively by quantum Monte Carlo methods
\cite{buechlerPRL07,astraPRL07,dipolePRL09,filinovPRL10,filinovPRA12} for a wide range of
dimensionless densities $nr_0^2$, including high densities where
the dipole-dipole repulsion leads to such strong in-plane correlations that the excitation
spectrum exhibits a roton similar to the roton in superfluid $^4$He,
and even higher densities where the ground state
of the 2D DBG is a triangular crystal.  Such large $nr_0^2$ may soon be in
experimental reach because $r_0$ can be very large, as mentioned above.
For the 2D DBG with {\em tilted} polarization a stripe
phase has been predicted recently for sufficiently large tilt angle~\cite{maciaPRA11,maciaPRL12}.
Also the more complicated case of a quasi-2D layer was studied, i.e. of a DBG in a
one-dimensional trap $U_{ext}(z)$.  The finite extent in
this direction allows pairs of particles to explore the anisotropy of the DDI.
Already by using the mean field approach of the Gross-Pitaevskii equation, ``rotonization''
of a quasi-2D layer of a polarized DBG was found to occur if the strength of the DDI surpasses
a critical value with respect to the short-range repulsion\cite{santosPRL03,wilsonPRL08}.
The roton in this case is not a signature of the repulsive correlations as
in the 2D limit for high densities, but a signature of the attractive correlations
for head-to-tail configurations of pairs of dipoles.  We have shown this
conclusively in Ref.~\onlinecite{hufnaglPRL11}, where even a cross-over
between these two kinds of rotons was demonstrated by variation of the trap length.
While the effect of attractive correlations on the excitation spectrum can
be qualitatively described by mean field methods that optimize only the ground state
density by the Ritz' variational principle, repulsive correlations leading to
$^4$He-like rotons require the optimization of at least density and {\em pair}
density.  This is exactly what the family of hypernetted chain Euler-Lagrange (HNC-EL) methods
does.  The HNC-EL method was therefore used in Ref.~\onlinecite{hufnaglPRL11} in
order to investigate both kinds of rotons.  A summary of how
HNC-EL works is given in the next section, all details about the method can
be found in Ref.~\cite{KroQiaKoh}.

In this work we extend our previous investigation of
a single quasi-2D DBG layer~\cite{hufnaglJLTP10,hufnaglPRL11}
to a bilayer.  The coupling between layers via the long-ranged DDI and the possibility of
pairing of dipoles on different layers to form dimers (and more generally of $n$ bound
dipoles in $n$ layers) has been investigated previously~\cite{wangPRL07,trefzgerPRL09}.
Superfluidity of fermionic bilayers was studied in Ref.~\cite{baranovPRA11}.
The bound and scattering states of just two dipoles on different layers have been
studied~\cite{volosnievJPhysB11,armstrongEPL10} as well.

Our bilayer is realized by a one-dimensional double well potential 
\begin{equation}
U_{ext}(z_i)=A\left\{\cos\left(qz_i-\pi\right)+\lambda\cos\left(2qz_i-2\pi\right)\right\} .
\label{eq:uext}
\end{equation}
A DBG in this trap is homogenous and infinitely large in $x$-- and $y$--direction and finite
in the confinement direction $z$.  The dipole moments are aligned along the $z$--direction,
therefore the dipole-dipole interaction (DDI) potential takes the form 
\begin{equation}
V_{dd}(\qr_i,\qr_j)=\frac{C_{dd}}{4\pi}\frac {1-3\cos^2\vartheta_{ij}}{r_{ij}^3}
\end{equation}
where $\vartheta_{ij}$ is the angle between the dipoles $i$ and $j$ measured from the $z$--axis,
and $r_{ij}=|\qr_i-\qr_j|$. To stabilize the system against collapse \cite{lahayePRL08} we add a hard
core repulsion, that is modeled by a $r_{ij}^{-12}$ potential. The Hamiltonian describing this
many--body system, in the reduced length and energy units, $r_0=mC_{dd}/(4\pi\hbar^2)$ and
$\epsilon_0=\hbar^2/(mr_0^2)$ respectively, looks as follows:
\begin{eqnarray}
  H&=&-\frac{1}{2}\sum_{i=1}^N \nabla_i^2+\sum_{i=1}^N U_{ext}(z_i) \\
&+&\sum_{i<j}V_{dd}(\qr_i,\qr_j) + \sum_{i<j}\Big(\frac {\sigma}{r_{ij}}\Big)^{12}.\nonumber
\end{eqnarray}
Correspondingly, we give all values for energy, length, wave number, and density in units of
$\epsilon_0$, $r_0$, $r_0^{-1}$, and $r_0^{-2}$; hence all quantities in tables and
figures are dimensionless.

In previous work \cite{odellPRL03,santosPRL03,hufnaglJLTP10,hufnaglPRL11}
it was established that a translationally invariant single layer of a DBG
in a one dimensional harmonic trap can become unstable due to the attractive part of the interaction.
The pair distribution calculated in Ref.~\onlinecite{hufnaglPRL11} shows that this instability can
be understood as a dimerization, where two dipoles can form a bound state; such weakly bound dipoles
would not be stable and indeed experiments with $^{52}$Cr~\cite{lahayePRL08} and
$^{168}$Er~\cite{aikawaPRL12} observe a collapse of the BEC for traps that are too
wide in the polarization direction.  Similarly, two coupled layers can become unstable
not only due to the attractive interaction within a layer,
but also due to the attraction between dipoles in different layers.  This latter
``instability'' actually indicates the dimerization of dipoles in different layers.
As long as the barrier between the layers is high enough that
the two bound dipoles remain in their respective layer, such a dimerized phase would be stable.

We have studied the DBG in the trap potential~(\ref{eq:uext}) for various
potential parameters $A$, $q$, and $\lambda$ that control the barrier height between
the wells, their separation, and their individual width.  We changed the parameters
such that we can study the transition
from two broad, but well-separated layers to two thin, but close layers.  We thereby go from a limit
that is dominated by {\em intra}-layer attraction to a limit dominated by {\em inter}-layer
attraction.  Both limits are characterized by the appearance of a soft mode (roton) with a respective
typical parallel wave number $k_\|x={\rm O}(1)$.  In the first limit, $x=a_{\rm ho}$ is the
oscillator length of the approximately harmonic well felt by each layer;
in the second limit, $x=d$ is the distance between the layers.
The six combinations of potential parameters $A$ and $q$ that we used
are listed in the first two columns of table~\ref{TAB::pot};
$\lambda$ was fixed to $\lambda=0.3$.  The corresponding trap potentials are plotted
in the lower panel of Fig.~\ref{FIG::Uextrho}, where we scale it by $100/A$ in order
to show all six potentials in the same figure.

We note that, although the average of the DDI over the whole plane vanishes, two dipoles on
different two-dimensional planes separated by a distance $d$ always dimerize, i.e.
form a weakly bound state due to the attractive head-to-tail well of the DDI, regardless of the value
of $d$~\cite{simonAnnPhys76}.  Hence in the zero density limit, the ground state is always dimerized.
At finite density, our calculations in the 2D limit show that dimerization is suppressed by
many-body effects~\cite{unpublished}.  In other words, increasing the density in the two
layers stabilizes the monomer phase.

\section{Ground State}

Ground state properties of Bose gases can be calculated using various methods, such as
the Gross--Pitaevskii method~\cite{gross61,pitaevskii61,dalfovoRMP99}
and quantum Monte Carlo methods~\cite{ceperley95,boronatValencia98,sarsaJCP00,boninsegniPRE06}.
The Gross--Pitaevskii method is widely applied for dilute systems, where the correlations between
particles are sufficiently weak such that the interaction between them can be
approximated by an effective mean potential felt by each particle.
Quantum Monte Carlo on the other hand is also suited for strongly interacting systems,
but is computationally demanding. 
For our calculations of the ground state properties we use the hypernetted--chain Euler
Lagrange (HNC-EL) method, which is a variational method suitable for strongly correlated
systems \cite{KroQiaKoh}, but with lower computational demands than QMC.  
Starting point is a Jastrow--Feenberg ansatz for the many-body wavefunction
\begin{equation}
\psi_0(\qr_1,\dots,\qr_N)=\exp\bigg[\frac 1 2 \sum_i u_1(\qr_i)+\frac 12\sum_{i<j}u_2(\qr_i,\qr_j)\bigg]
\label{eq:jastrow}
\end{equation}
which is optimized by solving the Euler-Lagrange equations numerically 
$$
  \frac{\delta\left<H\right>}{\delta \rho(\qr)}=0\,,\qquad
  \frac{\delta \left<H\right>}{\delta g(\qr,\qr^\prime)}=0\,.
$$
Here $\rho(\qr)\equiv\rho_1(\qr)$ is the one-body density and
$g(\qr_1,\qr_2)=\rho_2(\qr_1,\qr_2)\left(\rho(\qr_1)\rho(\qr_2)\right)^{-1}$ is
the pair distribution function.  $\rho_1(\qr)$ and $\rho_2(\qr_1,\qr_2)$ are special cases
of the $n$-body density reduced from the full $N$-body probablity of the (normalized) full
wave function $\psi_0(\qr_1,\dots,\qr_N)$
\begin{eqnarray*}
\rho_n(\qr_1,\dots,\qr_n) \equiv&& \\
{N!\over (N-n)!}&&\int\! d^3r_{n+1}\dots d^3r_N\ |\psi_0(\qr_1,\dots,\qr_N)|^2
\end{eqnarray*}

Due to translational invariance in $x$ and $y$-direction, for the present 
layer geometry all two-body functions, such as the pair distribution function,
depend on three variables: the modulus of the projection of $\qr\equiv\qr_1-\qr_2$ on the plane,
$r_\|=\sqrt{(x_1-x_2)^2+(y_1-y_2)^2}$,
and the two $z$-components $z_1$ and $z_2$.  Hence we effectively have a pair distribution
function $g(z_1,z_2,r_\|)$.

\begin{table}[t!]
\begin{center}
    \begin{tabular}{ | c |c| c | c |c|}
\hline
   $q$  &  $A$ & $\Delta E$ & $\mu$ \\
    \hline \hline
$2.70$ & $2\cdot10^{4}$ &  $8.4\cdot10^{-1}$  & $15.86$\\ \hline
$2.00$ & $2\cdot10^{4}$ &  $5.6\cdot10^{-2}$  & $16.41$\\ \hline
$1.00$ & $2\cdot10^{4}$ &  $1.9\cdot10^{-6}$  & $16.05$\\ \hline
$0.50$ & $2\cdot10^{3}$ &  $3.6\cdot10^{-4}$  & $10.09$\\ \hline
$0.20$ & $1\cdot10^{3}$ &  $8.4\cdot10^{-9}$  &  $4.69$\\ \hline
$0.16$ & $5\cdot10^{2}$ &  $5.9\cdot10^{-8}$  &  $3.04$\\ 
    \hline
    \end{tabular}
\label{TAB::pot}
\caption{Table containing the potential parameters $A$ and $q$, the corresponding tunnel splitting
for a single particle and the chemical potential $\mu$ of the many-body system.}
\end{center}
\end{table}

All calculations are
done for a total area density of $nr_0^2=1$, i.e. $nr_0^2=1/2$ for each of the two layers.
The HNC equations can be formulated in terms of Mayer cluster diagrams known from
classical statistical mechanics~\cite{hansen}.  The exact solution of the HNC equations would
require the calculation of a class of diagrams called elementary diagrams that
cannot be summed exactly.  Elementary diagrams are especially important at high densities,
while they can be neglected a lower densities.  Furthermore, we note that we use
a Jastrow--Feenberg ansatz that does not extend to triplet correlations, $u_3(\qr_i,\qr_j,\qr_k)$.
From $^4$He we know that both elementary diagrams and triplet correlations are
important for quantitative agreement with experiment and quantum Monte Carlo simulations~\cite{Kro86}.
For the area density used in this work, we have checked the influence of the elementary diagrams
and the triplet correlations in the 2D limit.  In this limit correlations are stronger than
for quasi-2D geometries, so the 2D limit gives a conservative estimate of their importance.
We found that they improve the accuracy of the static structure function $S(k)$ (see below)
by less then 2\%.  Therefore we neglect elementary diagrams and triplet correlations. 
In Ref.~\cite{hufnaglJLTP10} we compared results for a 2D system of aligned dipoles to QMC 
calculations~\cite{astraPRL07} for the density $\rho_0=2$. 
The agreement between the HNC-EL results and the QMC calculations for the static structure function
is very good. The total energy for both calculations differs by about 2.8\%, if elementary diagrams and 
triplet correlations are neglected. For the even smaller area density considered here,
we expect the accuracy to be even better.

\begin{figure}[t!]
\begin{center}
\includegraphics*[width=0.495\textwidth]{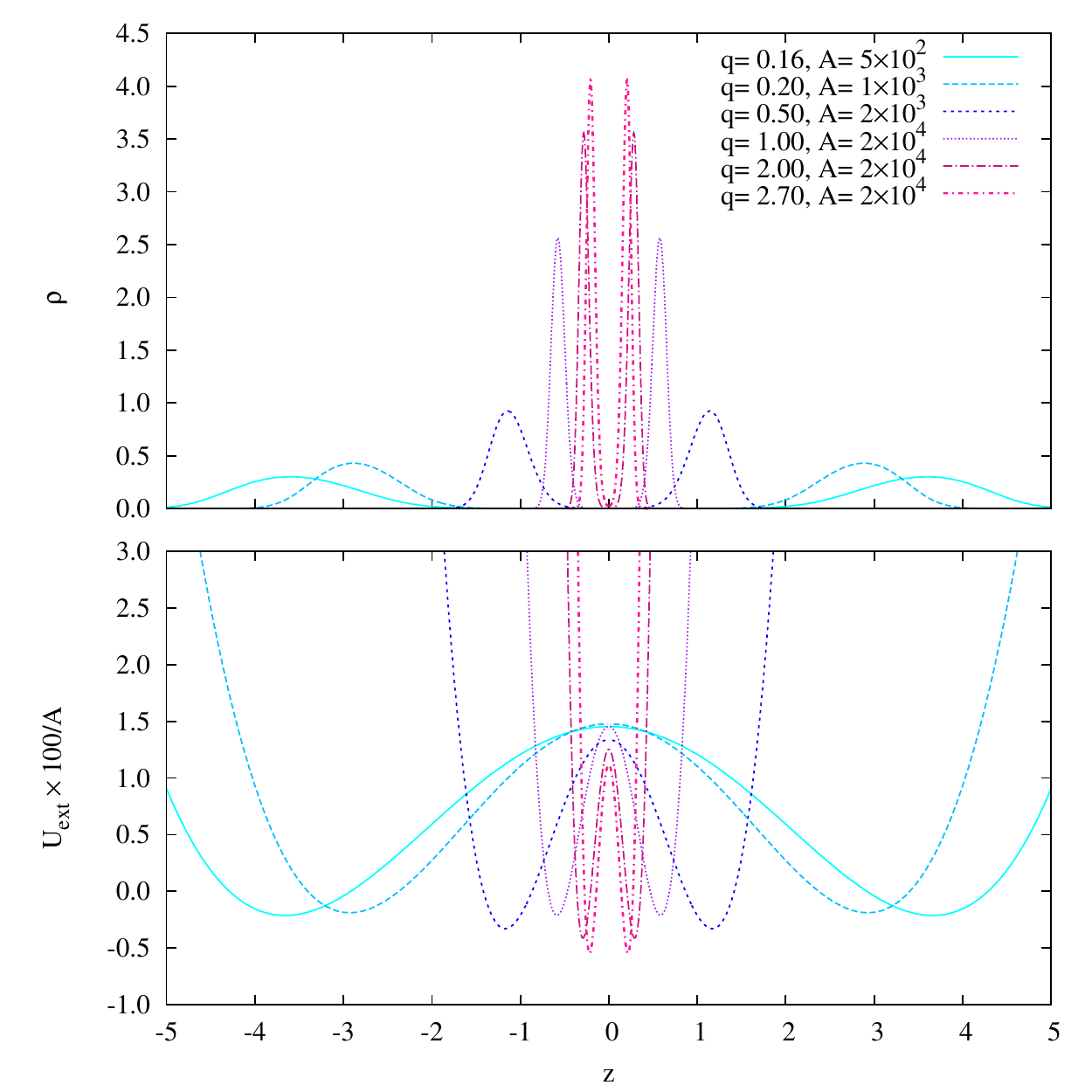}
\end{center}
\caption{ (Color online)
  The density profile $\rho(z)$ is shown in the upper panel and the trapping potential $U_{ext}(z)$
  according to eq.~(\ref{eq:uext}) is shown in the lower panel.
  The respective trap parameters $q$ and
  $A$ are listed in the upper panel.  $U_{ext}(z)$ was scaled by a factor
  of $\frac{100}{A}$ in order to show all potentials in the same figure.
}
\label{FIG::Uextrho}
\end{figure}

In table \ref{TAB::pot} we show the six parameter combinations that we choose
for the trap potential (\ref{eq:uext}) along with
the tunnel splitting for a single particle and the
chemical potential $\mu$ of the many-body system.  $\mu$ is measured with respect to the
single-particle ground state energy, i.e. with respect to a non-interacting Bose gas in
the same trap potential.  As expected, the chemical potential increases
if we decrease the thickness of the individual layers, which is due to the intra-layer repulsion
of both the DDI and the short-range interaction.  With increasing parameter $q$, the two
trap wells are not only getting closer, but, with our choice of parameter combinations, also
the tunnel splitting increases.
As mentioned above we gradually move from thick layers that are widely separated to thin layers
that are close to each other, while keeping the total area density fixed at $nr_0^2=1$.
The results for the density profiles $\rho(z)$ of the DBG in the trap potentials
can be seen in the upper panel of Fig.~\ref{FIG::Uextrho}.
The corresponding trap potentials $U_{ext}(z)$ are shown in the lower panel
using the same line style and color (online).  $U_{ext}(z)$ is scaled by the inverse of
the trap parameter $A$
such that all six potentials can be shown using the same scale.  Each potential is
offset such that the ground state energy of a single particle is zero.

\begin{figure}[t!]
\begin{center}
\includegraphics*[width=0.495\textwidth]{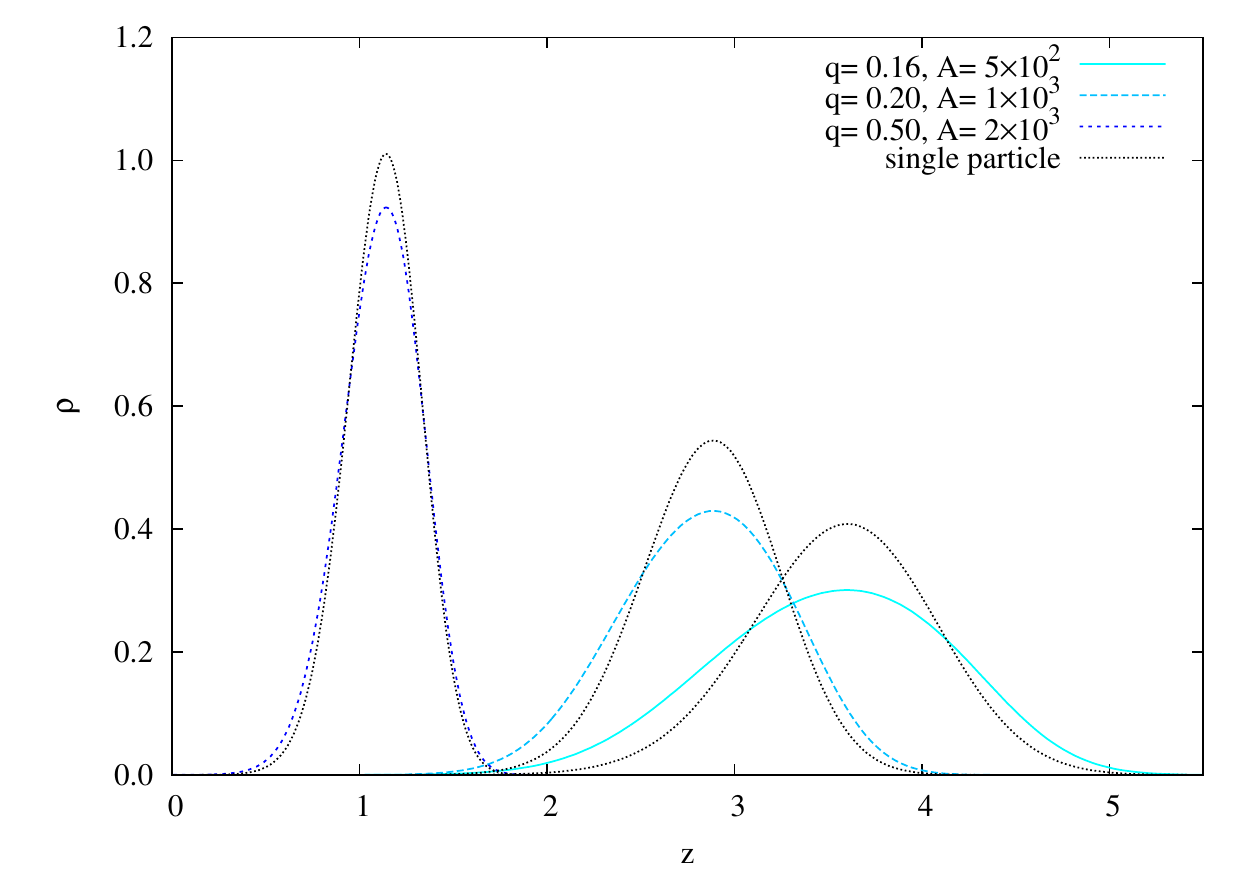}
\end{center}
\caption{(Color online)
  The density profile $\rho(z)$ is shown for the three wider traps
  and compared with the single-particle density $|\phi_0(z)|^2$ in
  the same traps (dotted lines).
}
\label{FIG::1body}
\end{figure}

In the limit of zero density or in the non-interacting limit,
the density profile is given by the square of the ground
state solution, $|\phi_0(z)|^2$, to the one-body Schr\"odinger equation
$H_1=-\frac{1}{2}\nabla^2+U_{ext}(z)$.  How closely the density $|\phi_0(z)|^2$ of the
one-body problem approximates the density $\rho(z)$ of the many-body problem depends
on the area density $n$ and the strength of the interactions, but also on the strength
of the trap potential.  For very tight confinement, the eigenenergies of $H_1$ 
above the ground state energy (or above the first two modes in case of a double well)
have energies so high that their contributions to the $N$-body ground state can
be neglected.  The weaker the confinement,
the more $|\phi_0(z)|^2$ and $\rho(z)$ will differ from each other.  This is indeed what
we find for the three weaker traps with $(q,A)=(0.50,2\cdot10^{3}),(0.20,1\cdot10^{3})$,
and $(0.16,5\cdot10^{2})$.  The comparison in Fig.~\ref{FIG::1body} between $|\phi_0(z)|^2$ and $\rho(z)$
shows that the interactions lead to a wider density $\rho(z)$, that does not
agree anymore with the one-body assumption $|\phi_0(z)|^2$.  For the three more
confined geometries, $|\phi_0(z)|^2$ and $\rho(z)$ are almost indistinguishable
(not shown in Fig.~\ref{FIG::1body}).  Note that this does not mean that a DDI in a tight trap
is well described by the one-body Hamiltonian $H_1$; the opposite is true,
in-plane correlations are stronger in a tight trap~\cite{hufnaglPRL11}.

\begin{figure}[t!]
\begin{center}
\includegraphics*[width=0.49\textwidth]{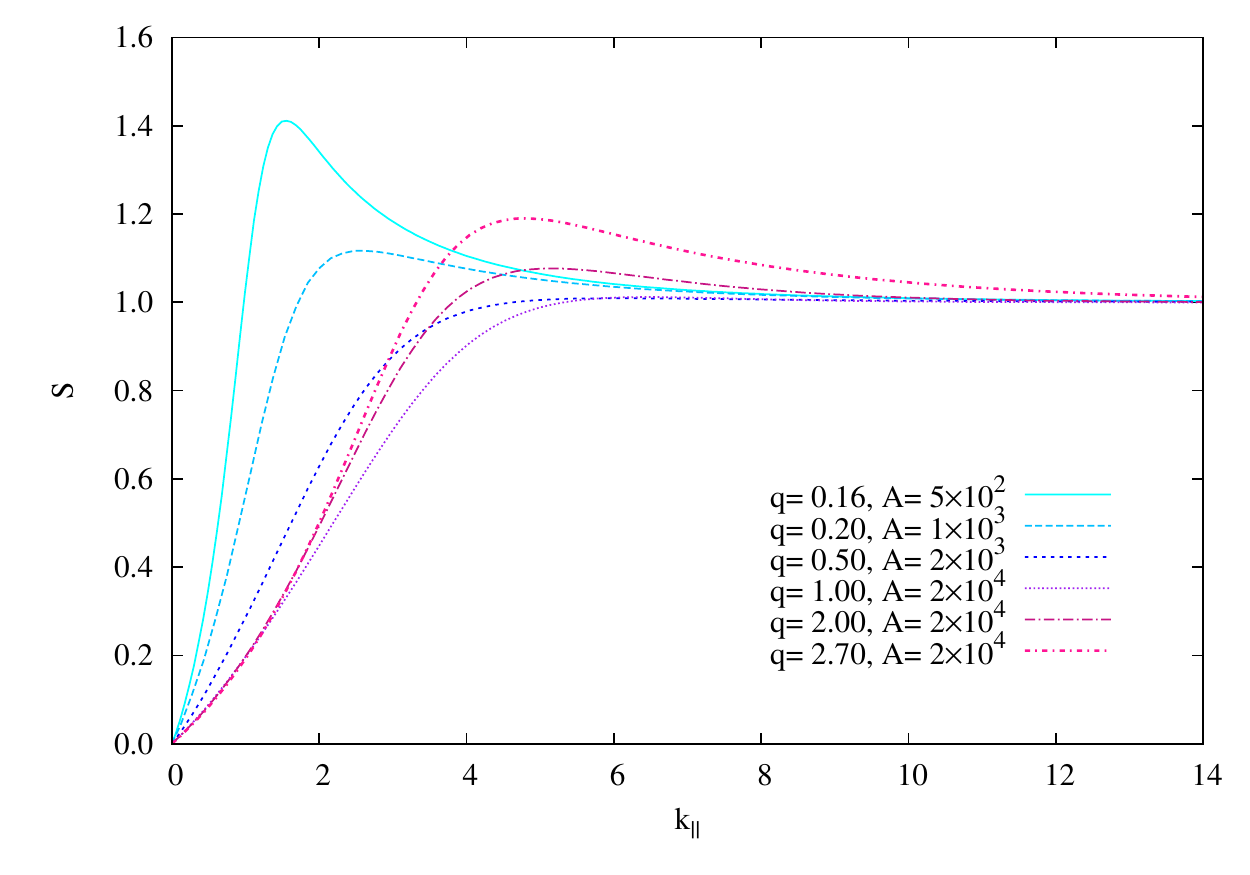}
\end{center}
\caption{(Color online)
Static structure function $S(k_\|)$  as a function of the parallel momentum $k_\|$ for the six different trapping potentials.
}
\label{FIG::S}
\end{figure}

For the layer geometry we define a static structure function $S(k_\|)$ as
\begin{eqnarray}
  S(k_\|) &=& 1 + \int d^3rd^3r'\, e^{i\qk_\|(\qr-\qr')} [g(\qr,\qr')-1]
\label{eq:S}\\
&=& 1 + \int dzdz' d^2r_\|\, e^{i\qk_\|\qr_\|} [g(z,z',r_\|)-1] \nonumber
\end{eqnarray}
where $k_\|$ is the parallel wave number,
$g(\qr,\qr')$ is the pair distribution function introduced above and $\qk_\|$ is any
wave vector in the $xy$-plane with wave number $k_\|$.  Note that $S(k_\|)$ depends
only on $k_\|$, while we integrate over the $z$ and $z'$ dependence of $g(\qr,\qr')$.
In Fig. \ref{FIG::S} we show the static structure function $S(k_\|)$ as a function of
$k_\|$ for the six traps studied here.  As we go from well-separated thick layers to close thin layers
we observe a peak in $S(k_\|)$ in both limits, whereas the peak vanishes in between.
It is natural to assume that
for wide layers the peak is caused by correlations due to the intra--layer
attraction of the dipoles whereas at for a small layer distance it is caused by correlations
due to the inter--layer attraction of the dipoles.  However, $S(k_\|)$ does not contain
enough information to distinguish between the these two mechanisms. 

\begin{figure}[t!]
\begin{center}
\includegraphics*[width=0.495\textwidth]{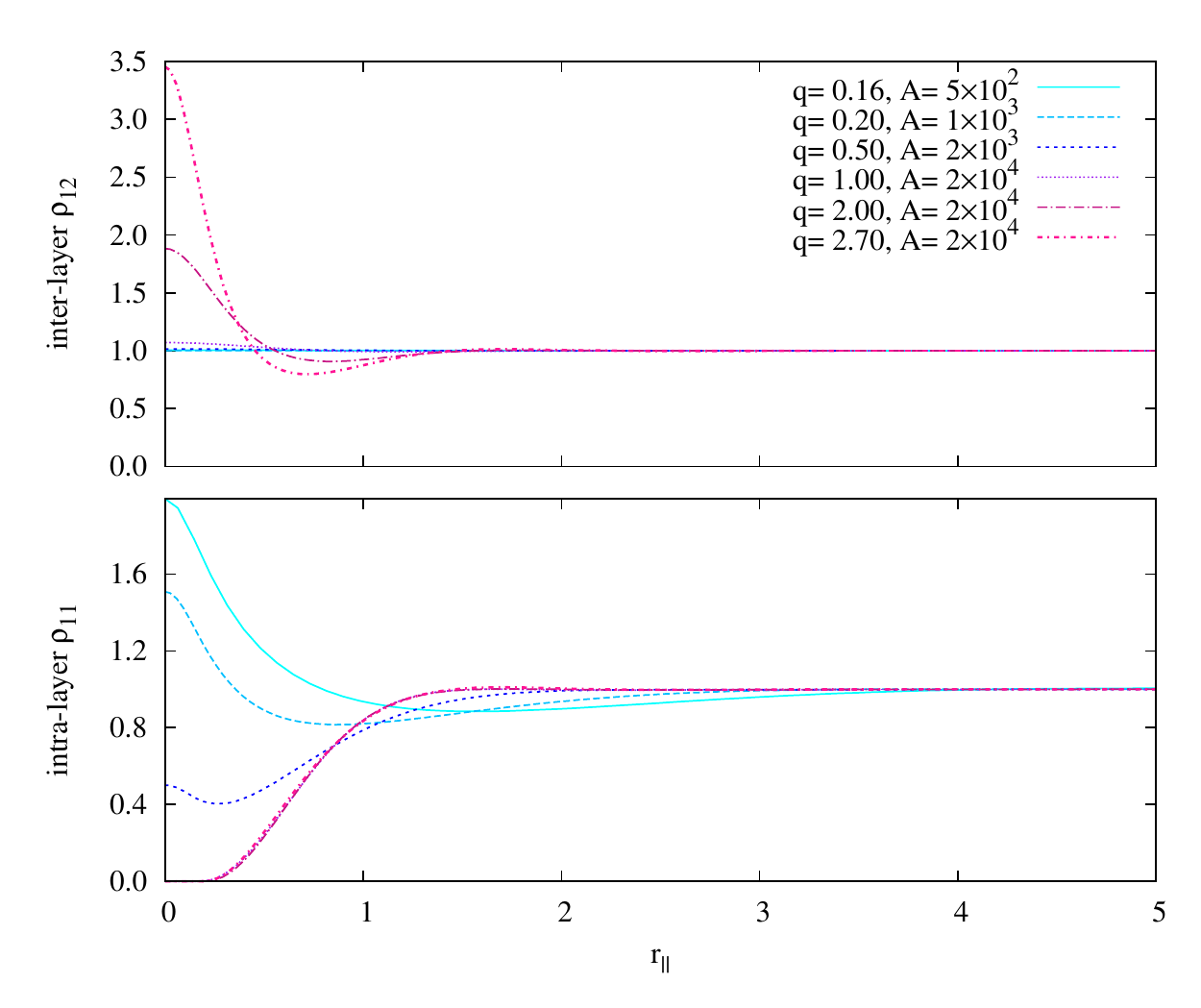}
\end{center}
\caption{(Color online)
Integrated two-body density (see text) for particles in the same layer $\rho_{11}$ (\emph{bottom panel}), and particles in different layers $\rho_{12}$ (\emph{top panel})
}
\label{FIG::rho2}
\end{figure}

In order to gain
information about intra- and inter-layer correlations, we define the partially
integrated pair densities $\rho_{11}(r_\|)$ and $\rho_{12}(r_\|)$,
\begin{eqnarray}
 \rho_{11}(r_\|)&=&\frac{4}{\rho_0^2}\int_{0}^{\infty}\int_{0}^{\infty} dzdz^\prime\,\rho_2(r_\|,z,z^\prime)
\label{eq:rho11}\\
 \rho_{12}(r_\|)&=&\frac{4}{\rho_0^2}\int_{-\infty}^{0}\int_{0}^{\infty} dzdz^\prime\,\rho_2(r_\|,z,z^\prime)
\label{eq:rho12}
\end{eqnarray}
The prefactors are chosen such that $\rho_{ij}(r_\|)\to 1$ for $r_\|\to\infty$, thus
$\rho_{11}(r_\|)$ and $\rho_{12}(r_\|)$ can be regarded as intra- and inter-layer pair
distribution functions.  They are the normalized probabilities to find two dipoles in the same layer
and in opposite layers at a parallel distance $r_\|$, respectively, regardless of
their $z$-coordinate within the layer.  $\rho_{11}(r_\|)$ and $\rho_{12}(r_\|)$ are shown
in the lower and upper panel of Fig.~\ref{FIG::rho2} for all six traps.
For wide, but well-separated layers there are strong intra-layer correlations at $r_\|=0$, whereas the
inter-layer correlations are vanishingly small. This means that two particles in the same layer
have a very high probability for head-to-tail configurations, with no parallel separation.
As we decrease the thickness of each layer, these intra-layer correlations vanish.  At the same
time we decrease the distance between layers, thereby increasing the inter-layer correlations.
For the smallest distance, two particles in {\em different} layers are strongly correlated and have a high
probability for head-to-tail configurations. Since the layer is thin, particles in the same layer
have a vanishing probability for zero parallel separation because of the DDI and the short-range repulsion.
In both limits of two independent wide traps and two close narrow traps the
respective strong positive correlations of $\rho_{11}$ and $\rho_{12}$ suggest a tendency
towards dimer formation, where two dipoles align head-to-tail either within a layer or across two layers. 

What happens, if we would drive the system to even larger correlation peaks in
$\rho_{11}(r_\|)$ or $\rho_{12}(r_\|)$?
The instability with respect to dimerization manifests itself as a numerical
instability of the HNC-EL equations.  Unlike other approximations, the
HNC-EL equations have the benefit that they do not produce a solution,
if a ground state of an assumed variational
form does not exist.  In the present case, the Jastrow-Feenberg ansatz
(\ref{eq:jastrow}) does not allow for the dimerization that our above analysis of
intra- and inter-layer pair distributions clearly suggests.  Since the ground state we try to
compute does not exist, our iterative procedure to solve the HNC-EL equations does
not converge.  In order to actually compute the properties of the dimerized
phase, one would have to optimize a variational ansatz that is flexible enough to
allow dimerization, or alternatively perform quantum Monte Carlo simulations, as e.g.\/
in Ref.~\cite{maciaPRL12}.

\section{Excitations}

\subsection{Bijl-Feynman modes}

\begin{figure}[t!]
\begin{center}
\includegraphics*[width=0.495\textwidth]{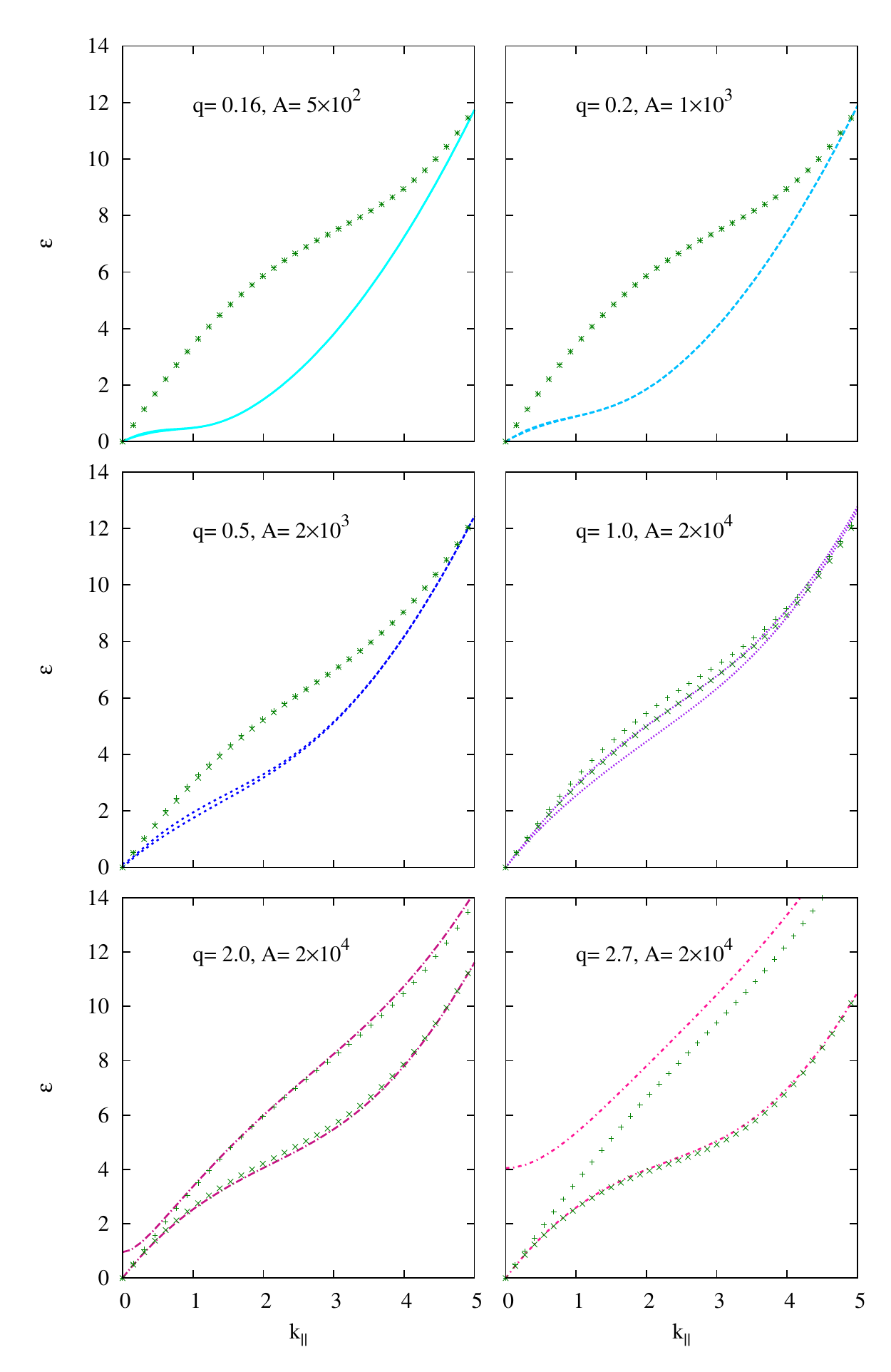}
\end{center}
\caption{(Color online)
Energy of the first and the second excitation mode in Bijl--Feynman approximation for different distances of the layers (lines), in comparison to the energies obtained for two 2D layers (dots).
}
\label{FIG::fey}
\end{figure}

\begin{figure}[t!]
\begin{center}
\includegraphics*[width=0.495\textwidth]{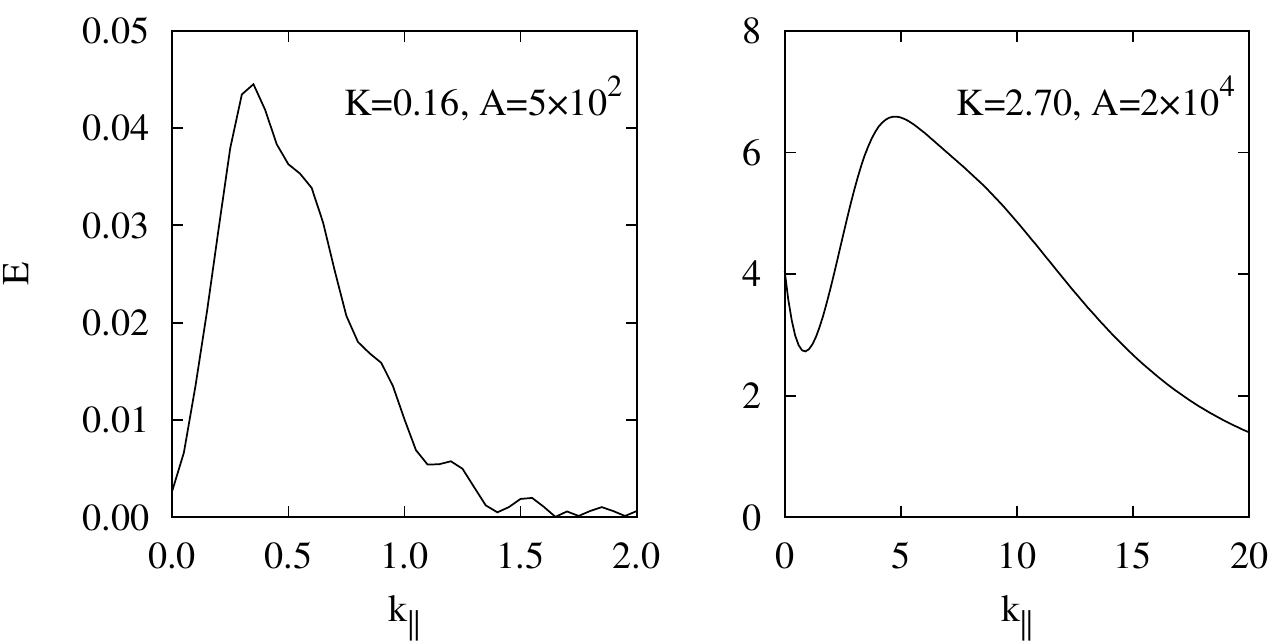}
\end{center}
\caption{Difference $\Delta\epsilon(k_\|)$ between
the energies of the two lowest excitation modes in Bijl-Feynman approximation.
}
\label{FIG:feyndiff}
\end{figure}

Owing to the translational invariance, excitations can be characterized by a parallel (i.e. in-plane)
wave number $k_\|$.  For a given $k_\|$ there are in principle an infinite number of
excitations, that are indexed by a perpendicular quantum number $n\in {\bf N}$ associated
with out-of-plane motion.  Especially for narrow double-well traps these modes have a much
higher energy than the lowest two modes in the interesting regime of wave numbers
$k_\|$, therefore we restrict our discussion to the two lowest modes and the
appearance of a soft mode.
In Fig~\ref{FIG::fey} we show the first and the second excitation mode, $\epsilon_1(k_\|)$
and $\epsilon_2(k_\|)$ in Bijl-Feynman approximation as a function of $k_\|$
for the six layer geometries for which we studied the ground state above.
The first and the second excitation mode are almost degenerate for a large distance between layers
and considerably split for a small distance.
Two completely independent DBG layers would of course have two-fold degenerate
excitation energies.  Even for the most separated
layers, the Feynman dispersion is not truly degenerate, but split for small
wave numbers.  In order to illustrate this, we show the energy difference
$\Delta\epsilon(k_\|)=\epsilon_2(k_\|)-\epsilon_1(k_\|)$ between the
two lowest Feynman energies as a function of $k_\|$ in Fig.~\ref{FIG:feyndiff} for the two traps
closest to instability.  The lifting of the degeneracy can
only be due to the DDI that is long ranged and hence couples
even well separated layers.  We can estimate the typical range of parallel wave numbers
for which excitations are most strongly affected.  We assume a circular density wave $\sim J_0(kr)$,
of wave number $k$, in one layer.  A particle at $r=0$ in the other layer will feel
a particularly strong dipole force if $k$ is such that $J_0(kr_0)=0$ where $r_0$ is the
radius where the DDI changes from attractive to repulsive. $r_0$ is
given by $r_0=d \tan\theta$, where $\theta$ is the angle of the attractive cone of the
dipole interaction, $\cos\theta=1/\sqrt{3}$, which gives $r_0=d\sqrt{2}$.
From this we get an estimate for the wave vector $k$ at which we observe the strongest
dipole coupling, which is $k=2.4048/(d\sqrt{2})$.
If we estimate $d$ as the distance between the two maxima of the density
profiles shown in the top panel of Fig.~\ref{FIG::Uextrho}, we obtain $k\approx 4$ and $k\approx 0.24$
for the closest and most separated layers, respectively.
This simple estimate agrees reasonably well with the
maximum energy splitting of the Feynman spectrum at $k=5$ and $k=0.35$ in Fig.~\ref{FIG:feyndiff}.
Note that for $k=0$ the DDI averages out, leading to zero DDI-induced
splitting for $k\to 0$, which is what we observe for well-separated layers.
For the closest layers the splitting at $k=0$ is large, however, which is
caused by our short-range repulsion model $(\sigma/r)^{12}$ which at such small $d$ can
be felt between different layers.  One could decrease $\sigma$ without compromising the
stability against intra-layer dimerization, but we preferred to tune only the external trapping
potential while keeping the interaction parameters fixed.  Furthermore, the tunnel splitting
is not small anymore for the closest layers, adding to the splitting caused by
the short-range repulsion.

In order to test our conclusions regarding inter-layer coupling for two layers of finite thickness, we also
performed calculations for the limit of two 2D layers. In this case the interaction within the same layer
is purely repulsive $\sim r^{-3}$ and the attractive part of the interaction is completely missing.
Positive correlations are possible only for the inter-layer pair distribution $\rho_{11}(r_\|)$.
The 2D results for the two lowest Bijl-Feynman energy dispersions are shown as symbols
in Fig.~\ref{FIG::fey}.
For the wide layers that are far apart, the quasi-2D and 2D results differ substantially (top left panel),
which demonstrates that the bending of the dispersion towards forming
a roton is an intra-layer effect.  As we make each layer narrower, the quasi-2D and 2D results
become almost identical. This means that the intra--layer attraction plays less of a role and
the roton formation is truly an inter--layer effect.  Note that in the 2D limit,
the splitting between the two lowest modes vanishes for $k_\|\to 0$ in agreement
with the above argument that the DDI averages out when integrated over the whole 2D plane.

\subsection{Dynamic structure function from CBF-BW}

Calculations of the excitations in the 2D limit of single layers have shown
\cite{dipolePRL09} that the Feynman approximation is
adequate for the dispersion relation only at very low densities, but correlation
effects become more important as the density is increased and fluctuations of pair
correlations must be taken into account.
Pair correlation fluctuations are accounted for in the correlated basis
function - Brillouin-Wigner (CBF-BW) formalism~\cite{Surface2}.
The CBF-BW method not only improves the accuracy of the excitation energies,
it also describes damping via decay of collective modes into lower energy modes.
We will see that the DDI coupling between layers leads to even larger
deviations of qualitative features of the excitation spectra in
the Feynman approximation.

The CBF-BW method was adapted to layer/film geometries
in Ref.~\cite{Clements96} and applied to superfluid $^4$He films
\cite{Clements96,apajaPRL03,apajaPRB03} and recently to single layers of
a DBG~\cite{hufnaglPRL11}.  The CBF-BW method has been demonstrated to
yield excitation energies much closer to the experimental results than
the Feynman approximation, even for such a strongly correlated system as
$^4$He.  Further improvement had been achieved for bulk $^4$He by including
fluctuations of triplet correlations~\cite{campbellJLTP10}.  The added complexity,
however, precludes an application to inhomogeneous systems.

\begin{figure}[t!]
\includegraphics[width=0.499\textwidth]{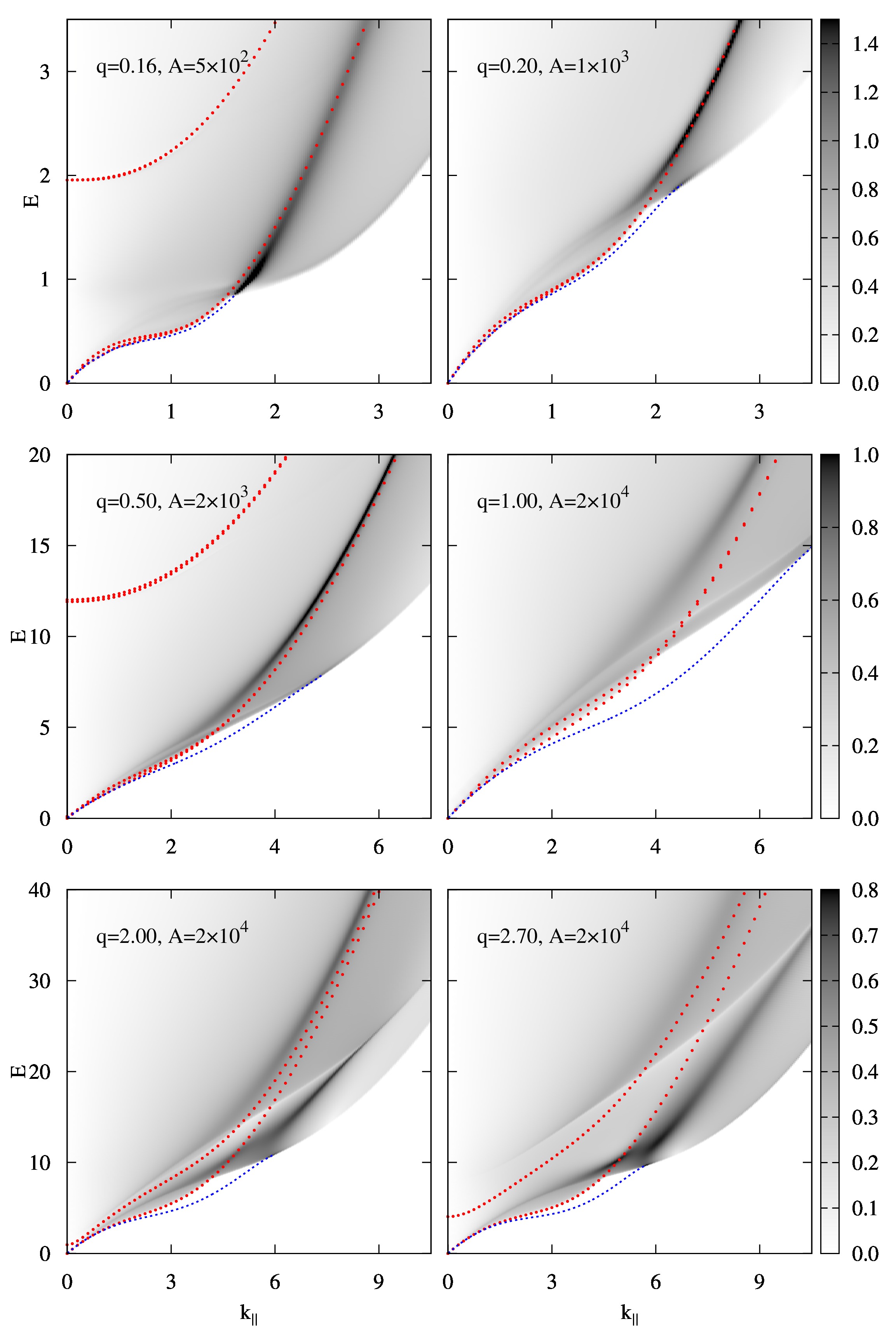}
\caption{(Color online)
$S(k_\|,E)$ is shown for six different traps, with the trap parameters $q$ and
$A$ given in each panel.  For better visibility of low-intensity features, we
map $S(k_\|,E)^{1/4}$ to a gray scale.  The full lines trace undamped peaks, the
dotted lines are the dispersion relations in Bijl-Feynman approximation.}
\label{FIG:skw2x3}
\end{figure}

The CBF-BW excitation energies are conveniently obtained by following the linear response
approach that yields the density-density response operator $\chi(\qr,\qr',E)$ and -- via
the fluctuation-dissipation theorem~\cite{Forster75} -- the
dynamic structure function $S(\qr,\qr',E)=-\Im m\chi(\qr,\qr',E)/\pi$,
where $E/\hbar$ is the frequency of a small external perturbation.
The derivation of $\chi(\qr,\qr',E)$ can be found in Ref.~\onlinecite{Clements96}.
If we project $S(\qr,\qr',E)$ onto plane waves
$$
  S(\qk,E) = \int d^3r d^3r'\, e^{i\qk(\qr-\qr')}S(\qr,\qr',E)
$$
we obtain the inelastic cross section for a perturbation imparting the momentum
$\hbar\qk$ to the system.
For a given $\qk$, a peak in $S(\qk,E)$ at an energy $E=\bar E$ indicates an excitation
of energy $\bar E$.  Peaks can have zero linewidth, if decay of an excitation
is kinematically forbidden, or finite linewidth otherwise.
Translation invariance in the $xy$-plane implies
that the projection $\qk_\|$ of $\qk$ on the $xy$-plane is a good quantum number.
A perturbation transferring a parallel momentum $\hbar\qk_\|$ and energy $E$
to the system probes the dispersion relation $\epsilon_n(k_\|)$
of the collective excitations, that we have calculated above in the simpler Feynman
approximation.  One might think that, since only the parallel component of $\qk$
matters for measuring the dispersion relation, we can restrict ourselves
to a parallel $\qk$, with a vanishing perpendicular component $\qk_\perp$.
However, the corresponding dynamic structure function $S(\qk_\|,E)$ probes
only excitation modes of {\em even} symmetry with respect to the $xy$-plane.
Since we are interested not just in the lowest (even) mode but also in the second (odd) mode,
we will show $S(\qk,E)$ also for wave vectors $\qk$ which have an angle $\theta$
with the $xy$-plane.  A purely perpendicular perturbation ($\qk_\|=0$)
could be implemented by fluctuations of the trapping potential (\ref{eq:uext}) itself,
but such a perturbation does not probe the dispersion relation.

\subsubsection{Parallel Momentum Transfer}

In Fig.\ref{FIG:skw2x3} we show $S(k,E)$ for the six different traps
shown in Fig.\ref{FIG::Uextrho}, the parameters given in table~\ref{TAB::pot} and
for wave vectors $\qk$ that are parallel to the $xy$-plane, i.e. $k_\perp=0$ and
hence $k_\|=k$.  $S(k,E)$ is represented in Fig.~\ref{FIG:skw2x3} by
mapping $S(k,E)^{1/4}$ to a gray scale.  The power of ${1\over 4}$ makes sure
that also broad, but low-intensity features can be seen well.  Full lines track peaks of
$S(k,E)$ of zero linewidth, i.e. which are proportional to a $\delta$-function.
The resulting line is an undamped dispersion relation.  For sufficiently large
wave number $k$, the dispersion relation merges with the gray area, where damping
by decay of an excitation into two lower-energy excitations
is kinematically possible (i.e. energy and momentum are conserved).
The Bijl-Feynman spectrum is shown as dotted lines
for comparison, including also higher modes.  For wider, well separated
traps, the Bijl-Feynman dispersion
agrees quite well with the CBF-BW result -- even for the widest trap
where a roton starts to form due to the intra-layer instability (top left panel).
Of course, the Bijl-Feynman approximation does not account for damping.
As we confine the two layers more strongly by increasing both trap potential
parameters $A$ and $q$, the dipole coupling between films leads to a splitting
of the Bijl-Feynman energies, as discussed above.  $S(k,E)$ has a much richer
structure that is poorly represented by the
Bijl-Feynman spectrum.  On the one hand the density increases as the trap tightens
(see Fig.\ref{FIG::Uextrho}), and the Bijl-Feynman approximation becomes worse
at higher density.  On the other hand, the DDI between layers
leads, in addition to a splitting of excitation energies, also to more
decay channels.

\begin{figure}[ht!]
\includegraphics[width=0.47\textwidth]{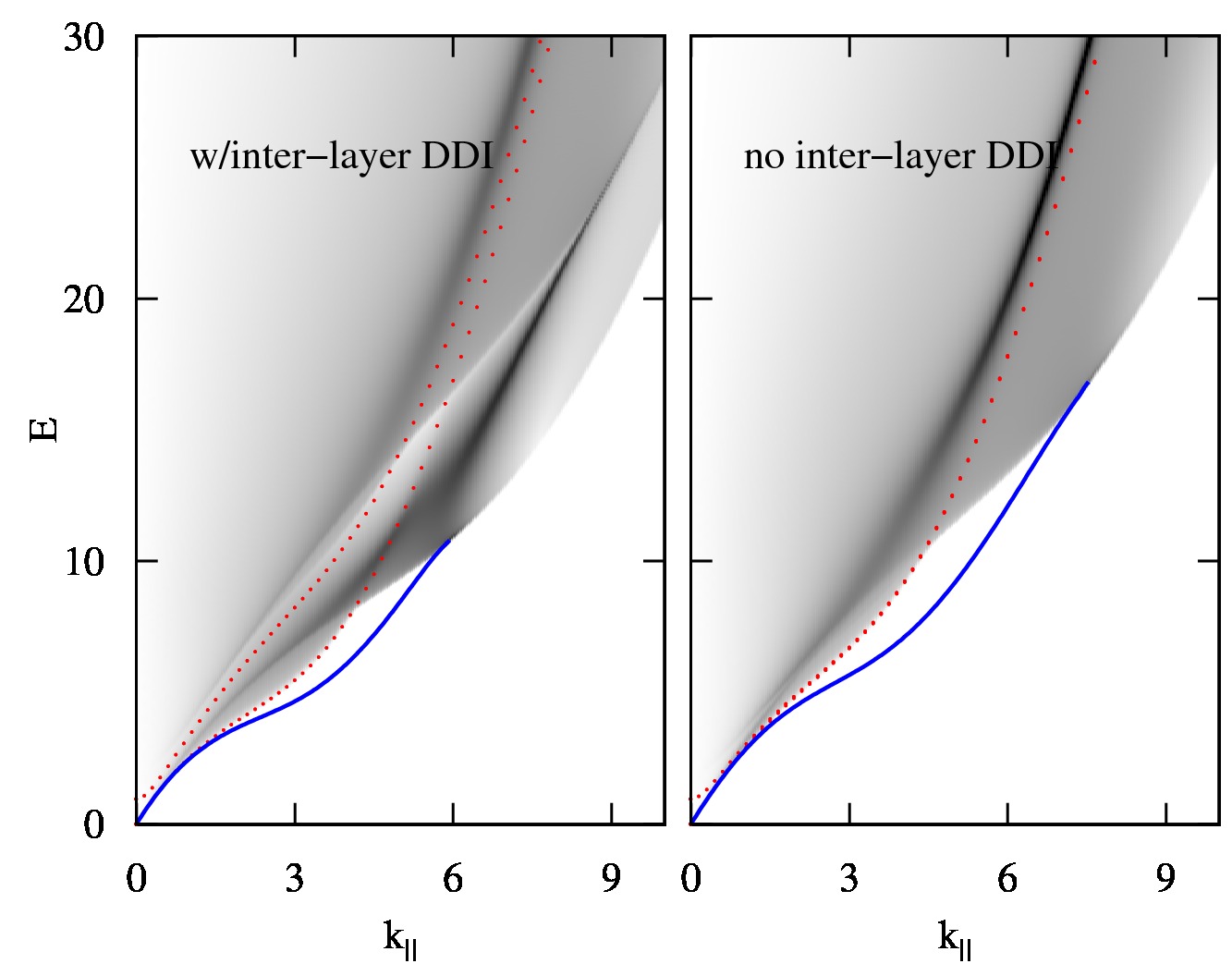}
\caption{(Color online)
The left panels shows $S(k_\|,E)$ for $A=2\cdot10^4$
and $q=2.0$ as in Fig.\ref{FIG:skw2x3}, i.e. with inter-layer DDI.  The right panel
shows the corresponding $S(k_\|,E)$ when the DDI between the layers is switched off.
}
\label{FIG::woint2}
\end{figure}

\begin{figure}[ht!]
\includegraphics[width=0.47\textwidth]{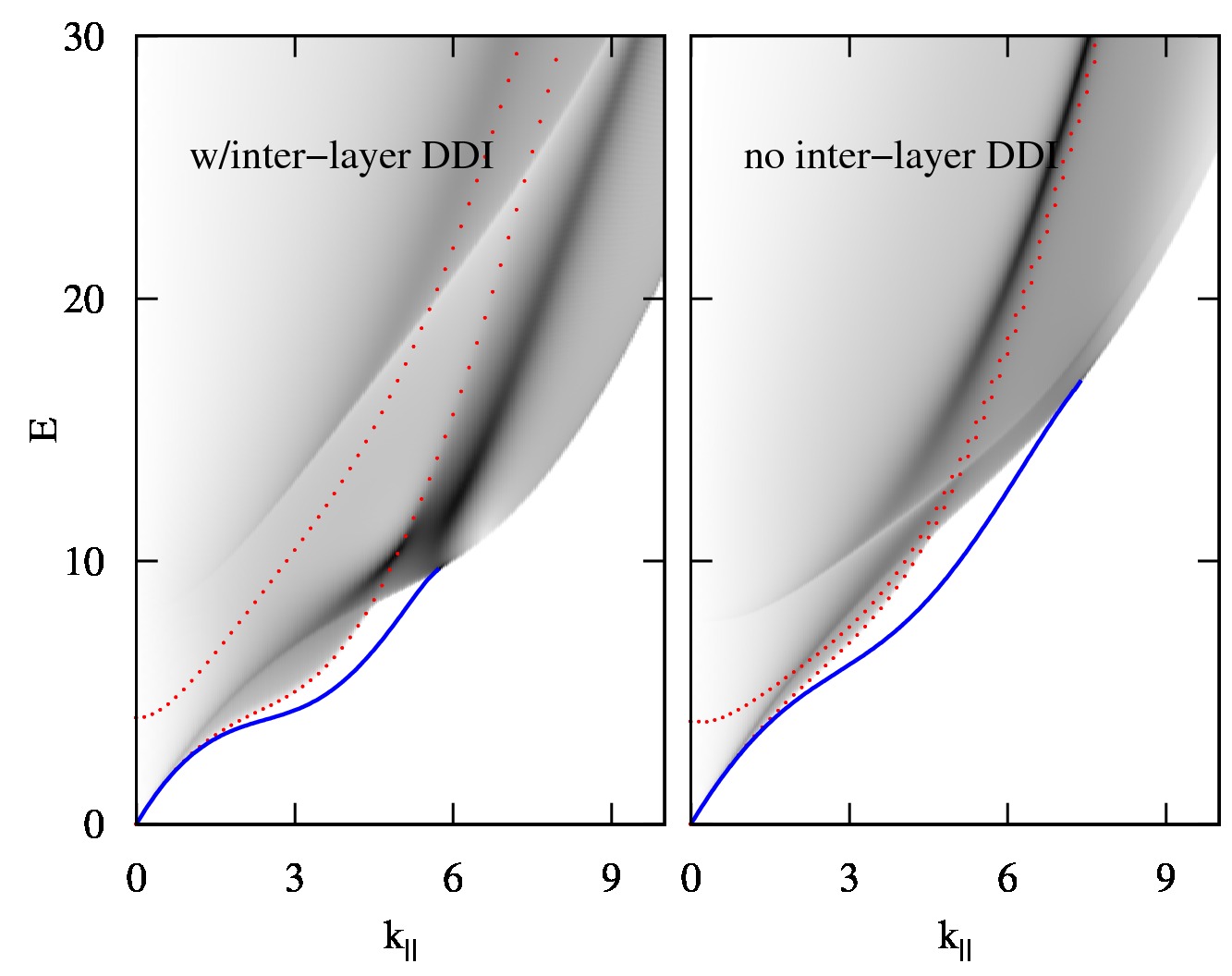}
\caption{(Color online)
Same as Fig.~\ref{FIG::woint2} for $A=2\cdot10^4$ and $q=2.7$.
}
\label{FIG::woint2.7}
\end{figure}

We demonstrate the importance of inter-layer DDI coupling by switching it
off for the two traps resulting in the closest layers ($A=2\cdot 10^4$ and $q=2.0; 2.7$).
This is simply achieved by setting $V_{dd}$ to zero if $z_1$ and $z_2$ have
opposite signs.  In Figs.~\ref{FIG::woint2} and \ref{FIG::woint2.7} we show
$S(k_\|,E)$ with the full DDI in the left panels and without inter-layer DDI
in the right panels.  For $q=2.0$ (Fig.~\ref{FIG::woint2}),
the lack of inter-layer DDI almost completely
decouples the two layers, leading to an almost degenerate Bijl-Feynman spectrum.
What we get is the dynamic structure function of a single layer, which has been
studied in Ref.~\onlinecite{hufnaglPRL11}.  The inter-layer DDI
leads to significant additional damping for higher energies, seen by the wider
peak in the energy regime where the dispersion becomes approximately quadratic.
Note that even without inter-layer DDI, there is a bend in the dispersion
for $q=2.0$, which shows it is not so much caused by inter-layer coupling,
but by the intra-layer repulsion of the DDI that, for much higher area density,
results in the type of roton studied in Ref.~\onlinecite{dipolePRL09} in the
2D limit.

\begin{figure}[ht!]
\includegraphics[width=0.47\textwidth]{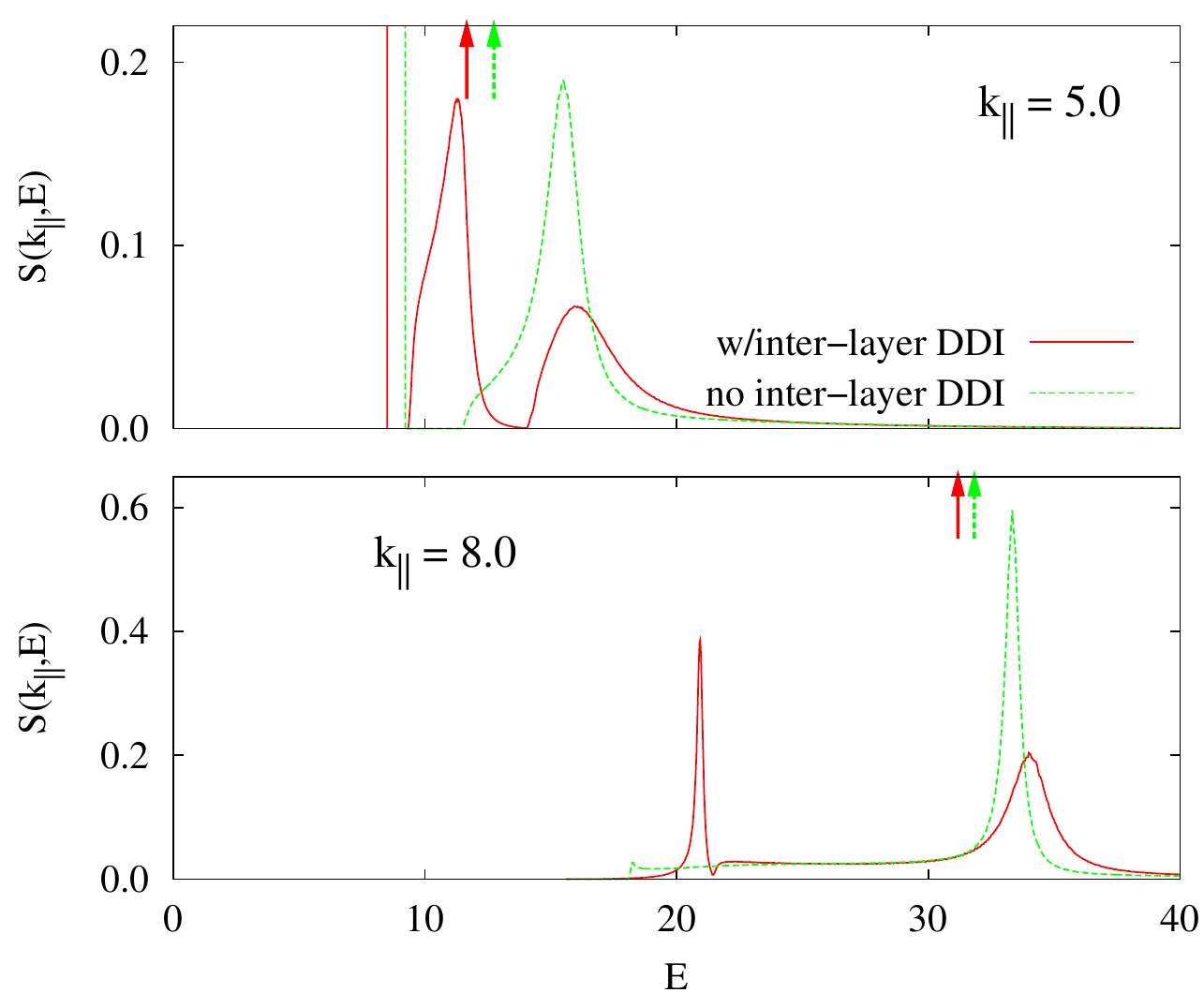}
\caption{(Color online)
A slice of $S(k_\|,E)$ with (full line) and without (dashed line)
inter-layer DDI, with $A=2\cdot10^4$ and $q=2.0$.  $k_\|=5.0$ in the upper panel
and $k_\|=8.0$ in the lower panel.  The vertical lines in the upper panel indicate
the respective energies of the undamped mode.  Arrows show the respective excitation
energies in Bijl-Feynman approximation.
}
\label{FIG::woint2-slice2}
\end{figure}

While a full $S(k_\|,E)$ map is necessary to track the dispersion relation,
the detailed line shapes of the various peaks are best seen by plotting
slices of $S(k_\|,E)$ for fixed values of $k_\|$.
The top panel of Fig.~\ref{FIG::woint2-slice2} shows, for trap parameters $A=2\cdot10^4$ and $q=2.0$,
a slice of $S(k_\|,E)$ at $k_\|=5.0$, which is slightly below the value of $k_\|$
where the sharp dispersion curve merges into the damping regime and thus becomes broad
(see full $S(k_\|,E)$ map in Fig.~\ref{FIG::woint2}).
The full line and dashed line are the results for $S(k_\|,E)$ with and without inter-layer DDI,
respectively.  The corresponding excitation energies in Bijl-Feynman approximation are indicated
by arrows.  The vertical lines are the undamped peaks
of the sharp dispersion.  We see that
$S(k_\|,E)$ has only one broadened peak without inter-layer DDI, while the inclusion of
the inter-layer DDI leads to two broadened peaks.  We stress again that for
parallel momentum transfer, $S(k,E)$ only probes the lower, even mode,
hence the two broad peaks are {\em not} due to the splitting of a degenerate
eigenmode ($S(k_\|,E)$ for non-parallel momentum transfer is presented below).

As $k_\|$ is increased further, the sharp peak loses more and more
spectral weight and eventually becomes damped.  This case is 
shown in the lower panel of Fig.~\ref{FIG::woint2-slice2}, where
$k_\|=8.0$ and the sharp dispersion has vanished for both the coupled and
uncoupled bilayer, see Fig.~\ref{FIG::woint2}.  Again there is only a single
peak without inter-layer DDI and two peaks with inter-layer DDI.
The lower peak is caused by the DDI coupling while the higher
one is only shifted slightly with respect to its position without DDI coupling.
Note that the DDI coupling approximately doubles the width of the
higher peak, hence reduces the lifetime of the associated excitation by about
a factor of two.  Thus, as one can expect, the dipole-dipole coupling between layers leads to
faster decay of excitations compared to uncoupled layers.

Finally, in Fig.~\ref{FIG::woint2.7} we compare $S(k,E)$ with and without inter-layer
DDI for even closer layers ($A=2\cdot10^4$ and $q=2.7$).
The bending is now significantly enhanced by the inter-layer DDI.
In CBF-BW approximation, the dispersion (blue line)
has a small slope at $k_\|=3$, i.e. the system is close to ``rotonization''.
Note that the residual splitting of the dispersion without
inter-layer DDI is due to tunneling and the short-range repulsion as mentioned above.

\subsubsection{Non-Parallel Momentum Transfer}

\begin{figure*}[t!]
\includegraphics[width=0.999\textwidth]{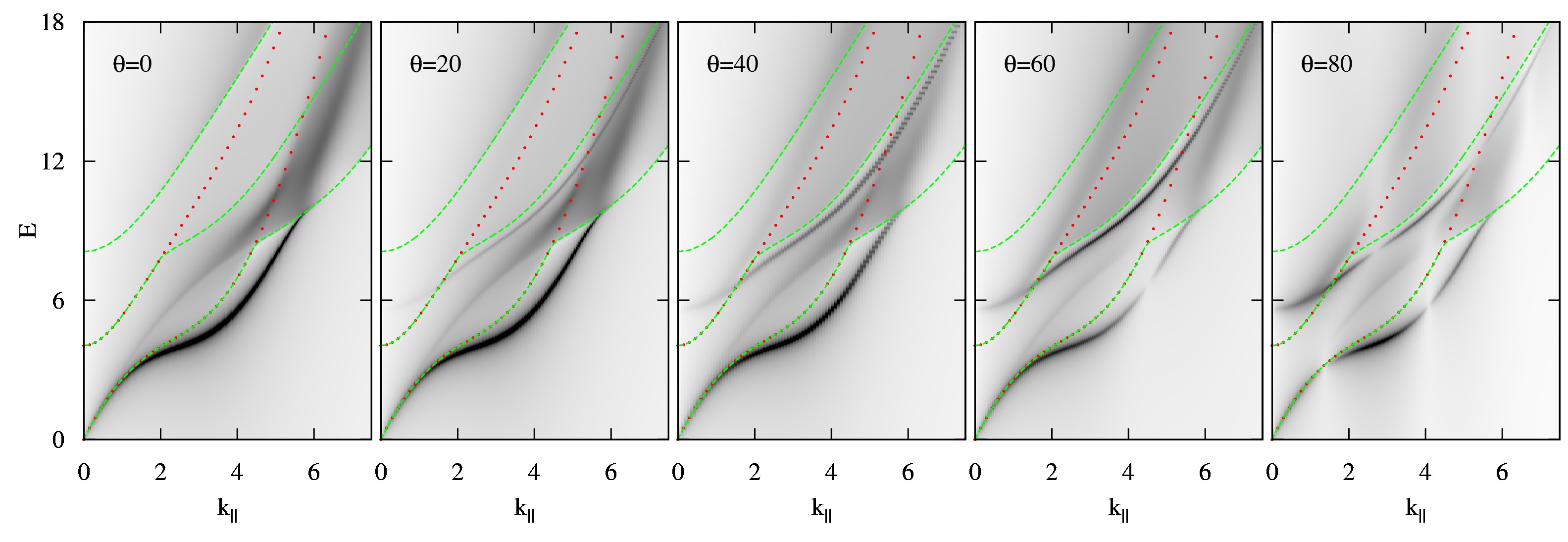}
\caption{(Color online)
$S(\qk,E)$ is shown as a function of $k_\|$ for different
angles $\theta$ of $\qk$ with respect to the plane of the bilayer.
The trap parameters are $A=2\cdot10^4$ and $q=2.7$ and
$S(\qk,E)$ was broadened by adding a small imaginary part to the energy, $\eta=0.1$.
As $\theta$ is increased, the second excitation modes becomes visible in $S(\qk,E)$.
The dots are the energies in Bijl-Feynman approximation, and the dashed lines are
the damping limits $E_{n,m}(k_\|)$ discussed in the text.
}
\label{FIG:theta2.7}
\end{figure*}

Parallel momentum transfer only probes those excitations
which are even with respect to inversion at the $z=0$-plane, because a perturbation
independent of $z$ is even and therefore cannot excite odd modes.  In order
to probe odd modes, we study $S(\qk,E)$ for wave vectors $\qk$ with an
arbitrary angle $\theta$ with respect to the $z=0$-plane.  Fig.~\ref{FIG:theta2.7}
shows $S(\qk,E)$ for $\theta=0;20;40;60;80$, for trap parameters
$A=2\cdot10^4$ and $q=2.7$.  We plot $S(\qk,E)$ as
a function of $k_\|$, not $|\qk|$, since only $k_\|$ is a good quantum number
that is meaningful for characterizing the excitation spectrum.  Unlike in
all previous figures of $S(\qk,E)$, we now add an artificial small imaginary part
$\eta=0.1$ to the energy $E$ which slightly broadens all features of
$S(\qk,E)$.  The rationale behind this broadening is that it makes
the spectral weight of peaks with zero intrinsic linewidth visible.

The case $\theta=0$ was shown already in Fig.\ref{FIG:skw2x3} (without artificial damping), and
is shown here again for better comparison with $\theta>0$.  For $\theta=0$
indeed only the lowest, even mode is visible.  As $\theta$ is
increased, a second mode becomes visible and gains weight.  For $\theta=60$, both modes
can be seen equally clear in $S(\qk,E)$ where they appear as a narrow dark trace.  Note that
the second, odd mode is damped for small $k_\|$.  This is very different from the low-$k_\|$
behavior of the lowest mode (sound mode) which is not damped because of its negative curvature.
The damping of the second mode can be seen as a broadening 
(in addition to the artificial broadening) for $k_\|\lesssim 1.8$.  We introduce the damping
limit $E_{n,m}(k_\|)$ which is the energy above which an excitation of parallel
wave number $k_\|$ can decay into two modes with perpendicular wave number $n$ and $m$.
$E_{n,m}(k_\|)$ is given by
$$
E_{n,m}(k_\|) = \min_{q_\|}[\epsilon_n(k_\|)+\epsilon_m(|\qq_\|-\qk_\||)]
$$
where, due to the limitations of the CBF-BW approximation,
$\epsilon_n(q_\|)$ are the excitation energies
in Bijl-Feynman approximation, not the excitation energies following from CBF-BW itself
(inclusion of triplet correlations has been shown for homogeneous systems to lead to
a self-consistent formulation of the self-energy, see Ref.~\cite{campbellJLTP10}).
Even modes can decay into combinations where $n+m$ is even and vice versa
for odd modes.  Since we are interested only in decays of the lowest two modes,
we obtain three decay limits, which fulfill $E_{1,1}(k_\|)<E_{1,2}(k_\|)<E_{2,2}(k_\|)$.
They are shown in Fig.\ref{FIG:theta2.7} as dashed lines.  The lowest
mode can decay into $(n,m)=(1,1)$ and $(2,2)$ and the second mode can decay
into $(n,m)=(1,2)$.  The effect of these respective limits are clearly seen in
Fig.\ref{FIG:theta2.7}.  Damping indeed sets in as the dispersion relation of the mode crosses
the damping limit with a symmetry appropriate for the mode.

Also visible in Fig.\ref{FIG:theta2.7} are interference patterns
that lead to a modulation of the intensity of $S(\qk,E)$ as $k_\|$ and thus
$k_\perp=k_\|\tan\theta$ is increased.  These are simply due to the perpendicular
wave number $k_\perp$ being in phase or out of phase with even or odd modes.
For example, a value of $k_\perp\approx \pi/d$, where $d$ is a measure
for the distance between the layers, leads to a cancellation of the
intensity for even modes, but to a maximal intensity for odd modes.

\section{Discussion and Conclusion}

In this work we generalized our previous studies~\cite{hufnaglPRL11}
of dipolar Bose gas layers from a single layer in a
harmonic trap to double-well traps which results in a bilayer geometry.
As in our previous work, the bilayer calculations are based on the HNC-EL
method for the many-body ground state and on the CBF-BW method for
excitations.  Dipolar bilayers have a richer structure than single layers
owing to the inter-layer dipole coupling.  The possibility
of head-to-tail pairing of two dipoles on different layers leads to similar rotonization
effects in the non-paired (monomer) phase as previously predicted in single layers.
We restricted ourselves to the calculation of
ground state correlations of the monomer phase, as well as its excitation spectrum,
including damping due to decay of excitations into two lower excitations.

We systematically varied the double-well trap parameters between two close, but
thin layers and two well separated, but wide layers, while keeping the total area density
fixed at a modest $nr_0^2=1$.  The two end points of the
range of trap parameters are marked by instabilities of the monomer phase.  Either if layers
are too close or if one layer is too wide, inter-layer or intra-layer dimerization occurs, respectively.
The latter kind of dimers are not stable and would quickly collapse via 3-body collisions,
but inter-layer dimers are stable, given a sufficiently high double well barrier.
The propensity to pairing was clearly seen in the monomer pair distribution functions $\rho_{12}(r_\|)$
or $\rho_{11}(r_\|)$, which are the normalized probabilities to find two dipoles in different or
the same layers, respectively, at a parallel distance $r_\|$.  We showed that, at $r_\|=0$,
$\rho_{12}(r_\|)$ grows a peak for small interlayer distance, while
$\rho_{11}(r_\|)$ grows one if each single layer is sufficiently wide.
In both cases, the peak of $\rho_{ij}(r_\|=0)$ is a precursor to the pairing of two dipoles
in head-to-tail orientation.

We presented calculation of the dynamic structure function $S(\qk,E)$ in the CBF-BW approximation.
$S(\qk,E)$ for parallel momentum transfer probes only even modes, where we are mostly
interested in the lowest one.  $S(\qk,E)$ typically consists of a lower,
undamped peak (that vanishes for higher $k_\|$) and two broad peaks that are due
the inter-layer DDI coupling (without it, there is only one broad peak), which also
enhances damping.  The double-peak structure is not to be confused with the more trivial
effect that each mode is split into two because the inter-layer DDI lifts its degeneracy.
The rich structure of $S(k_\|,E)$ is not captured by the simple Bijl-Feynman approximation which would
predict a single, undamped peak for the lowest mode.
The intra- and inter-layer instabilities of the monomer phase are characterized by a bending of the
dispersion relation of the lowest (intra-layer dimer) or lowest two (inter-layer dimer)
excitation modes.  This bending, that is less pronounced but still visible in the
Bijl-Feynman approximation, indicates ``rotonization'', which is well-studied
for single layers.  As in our previous work on single layers, we found that the iterative
procedure to solve the nonlinear set of HNC-EL equations becomes unstable as we approach
rotonization, i.e. as the dispersion relation starts to have a local minimum at finite
$k_\|$.  This leads to the conjecture
that the ground state is only metastable when the excitation spectrum exhibits
a roton, while the true ground state, i.e. the state of lowest energy is the
(intra- or inter-layer) dimerized phase.  For a proof of this conjecture, however, one would
need to compare our monomer results with results for the dimerized phase to find out
which state has the lowest energy.
Finally, we also presented results for non-parallel momentum transfer, i.e. where the angle
between $\qk$ and the plane of the layers is non-zero.  The dynamic
structure function depends on both the parallel and perpendicular components of $\qk$, $k_\|$ and
$k_\perp$. $k_\|$ still is a good quantum number, while the non-zero $k_\perp$ allows to
probe also odd modes, particularly the second excitation mode.  Showing $S(\qk,E)$ as function
of the parallel wave number, the second mode becomes clearly visible for e.g. an angle $\theta=60^\circ$.
Unlike the lowest mode, the second mode is damped for small $k_\|$ due to
decay into two excitations of lower energy.

An interesting topic are the correlations of the monomer phase and its excitations
generalized to $N$ layers.  The long-ranged DDI coupling between different layers
for example lifts an $N$-fold degeneracy of the excitation spectrum and opens many
possible decay channels for the resulting $N$ modes.  Another direction is the study of
``unbalanced'' bilayers where the two layers have different area densities, or bilayers
with different kinds of particles (e.g. different mass) on each layer.  If for example
the density in one layer is very low, the DDI interaction with the other layer would
constitute a very well controlled model of an impurity particle moving in one layer,
coupled to a bath of particles in the other layer.   The mechanism how an impurity
attains an effective mass could be investigated in a well-controlled fashion
over a much wider range of densities and
interaction strengths than in condensed matter systems.

\begin{acknowledgments}
We are grateful for discussions with Eckhard Krotscheck and Vesa Apaja.
We acknowledge financial support by the Austrian Science Fund FWF under grant No. \#23535
and by the National Science Foundation under Grant No. NSF PHY11-25915.
\end{acknowledgments}


%

\end{document}